\documentclass[aps, twocolumn,citeautoscript,floatfix]{revtex4-2}

\setcitestyle{super}
\usepackage{color}
\usepackage{amsmath}
\usepackage{array}
\usepackage{graphicx}
\usepackage{float}
\usepackage{booktabs}
\usepackage{algpseudocode}
\usepackage{algorithm2e}
\usepackage{physics}
\usepackage{hyperref}

\begin{document}
\RestyleAlgo{ruled}
\title{}
\author{Harper R. Grimsley}
\affiliation{Chemistry Department, Virginia Tech}
\author{George S. Barron}
\affiliation{Physics Department, Virginia Tech}
\author{Edwin Barnes}
\affiliation{Physics Department, Virginia Tech}
\author{Sophia E. Economou}
\affiliation{Physics Department, Virginia Tech}
\author{Nicholas J. Mayhall$^*$}
\affiliation{Chemistry Department, Virginia Tech}
\email{nmayhall@vt.edu}
    \date{\today}
\title{ADAPT-VQE is insensitive to rough parameter landscapes and barren plateaus}
\begin{abstract}
Variational quantum eigensolvers (VQEs) represent a powerful class of hybrid quantum-classical algorithms for computing molecular energies.  Various numerical issues exist for these methods, however, including barren plateaus and large numbers of local minima.  In this work, we consider Adaptive, Problem-Tailored (ADAPT)-VQE ans{\"a}tze, and examine how they are impacted by these local minima.  We find that while ADAPT-VQE does not remove local minima, the gradient-informed, one-operator-at-a-time circuit construction seems to accomplish two things:  First, it provides an initialization strategy that is dramatically better than random initialization, and which is applicable in situations where chemical intuition cannot help with initialization, i.e., when Hartree-Fock is a poor approximation to the ground state.  Second, even if an ADAPT-VQE iteration converges to a local trap at one step, it can still ``burrow'' toward the exact solution by adding more operators, which preferentially deepens the occupied trap.  
This same mechanism helps highlight a surprising feature of ADAPT-VQE: It should not suffer optimization problems due to ``barren plateaus.''
Even if barren plateaus appear in the parameter landscape, our analysis and simulations reveal that ADAPT-VQE avoids such regions by design.
\end{abstract}
\maketitle
\section{Introduction}
Quantum computers have long been viewed as a promising technology for quantum simulation.\cite{feynman_simulating_1982}  However, the limited capabilities of Noisy, Intermediate-Scale Quantum (NISQ) devices restrict the types of algorithms that can be implemented at present.  \cite{preskill_quantum_2018} 
While quantum phase estimation (QPE) provides a route to efficient molecular simulation,\cite{Aspuru-Guzik2005} the  presence of both noise and errors on NISQ devices 
make near-term implementation of large-scale phase estimation intractable.

In response to the intractability of QPE, the variational quantum eigensolver (VQE) was introduced by Peruzzo et. al. \cite{peruzzo_variational_2014} as a hybrid quantum-classical approach to finding approximate eigenvalues of a Hamiltonian, $\mathcal{H}$.  
In VQE, a quantum processor is used to apply a parameterized unitary transformation expressed as a quantum circuit (or even a direct pulse\cite{asthana_minimizing_2022,meitei_gate-free_2021,magann_pulses_2021}), $\mathcal{U}\left(\boldsymbol{\theta}\right)$, to some easily prepared reference state, $\ket{0}$.\cite{peruzzo_variational_2014,cao_quantum_2019,cerezo_variational_2021,tilly_variational_2021,fedorov_vqe_2022} 
The target Hamiltonian is then measured with the prepared state to obtain the energy as a function of circuit parameters:
\begin{equation}\label{eq:E}
E\left(\boldsymbol{\theta}\right) = \bra{0}\mathcal{U}^{\dagger}\left(\boldsymbol{\theta}\right)\mathcal{H}\mathcal{U}\left(\boldsymbol{\theta}\right)\ket{0}.
\end{equation}     
Using such quantum resources to prepare states and measure observables, a VQE will classically optimize $\boldsymbol\theta$ in order to minimize $E\left(\boldsymbol\theta\right)$.  The quality of the optimal energy for a given VQE is naturally dependent on the quality of the parameterization $\mathcal{U}\left(\boldsymbol\theta\right)$, but because unitary operators are norm-preserving, the energy in Eq.  (\ref{eq:E}) is variationally bounded from below by the ground-state energy of $\mathcal{H}$.  The main advantage of VQEs is relatively low circuit depth,\cite{peruzzo_variational_2014} avoiding the long, coherent evolutions of QPE.  \cite{kitaev_quantum_1995}  This makes VQEs more appealing in the absence of fault-tolerant quantum computers.  The circuit depth of a VQE is defined by the choice of $\mathcal{U}$, so that there is generally a trade-off between accuracy and circuit depth. 

An outstanding challenge with many VQE ans\"atze is that the cost function, Eq. (\ref{eq:E}), creates a rough parameter landscape full of local minima, complicating the parameter optimization.  Bittel and Kliesch have identified situations where there are so many far-from-optimal local minima that VQEs must be NP-hard in general.\cite{bittel_training_2021}  The problem of local minima can be ameliorated through overparametrization in both quantum optimal control\cite{riviello_searching_2015,asthana_minimizing_2022} and classical neural network settings.\cite{lopez-paz_easing_2018,du_gradient_2019}  This idea of overparametrization avoiding local minima has since been applied to VQEs:  Rivera-Dean et. al. used this philosophy by employing a neural network to distort their cost function landscape mid-VQE.\cite{rivera-dean_avoiding_2021}  This enabled them, in some cases, to escape from local minima.\footnote{The neural network temporarily adds additional ``weight'' parameters to the optimization.  Even when the neural network is then reset to the identity, a better set of parameters $\boldsymbol\theta$ was sometimes found for the undistorted cost function.}  Alternative strategies for avoiding local minima include collectively optimizing an ansatz for several Hamiltonians at the same time with a ``snake'' algorithm\cite{zhang_collective_2020} and a ``sweeping'' approach to energy minimization called Unitary Block Optimization.\cite{slattery_unitary_2021}

A recent theoretical analysis by Larocca et. al. suggests that quantum neural networks (of which VQEs are a special case) undergo a sort of phase transition where local minima cease to be a problem.\cite{larocca_theory_2021}  This transition tends to occur when the number of parameters surpasses the dimension of the associated ansatz's dynamical Lie algebra, or DLA.  The DLA for an ansatz of the form $\ket{\Psi} = e^{\theta_1 A_1}e^{\theta_2 A_2}\dots e^{\theta_M A_M}\ket{\phi_0}$ is defined as the span of the set of repeated commutators of $\{\hat{A}_i\}$. As the authors point out, their results imply that this desirable overparametrization is likely to be unachievable for ans{\"a}tze due to the exponential scaling of the DLA dimension with ansatz length.  Perhaps even more alarmingly, Wierichs et. al. were able to identify situations where adding additional parameters actually hurts the performance of gradient descent methods.\cite{wierichs_avoiding_2020}  

In addition to the problems with local traps, it has recently been recognized that VQEs might also become impossible to optimize (even to a local mininum) 
as the system size increases. 
For sufficiently flexible or expressive VQE ans\"atze \footnote{formal arguments have largely been restricted to 2-design structures}, it has been found that the energy landscape flattens (as quantified by the variance in the parameter gradients) exponentially fast as the system size increases.\cite{mcclean_barren_2018}
The exponential growth of these flat landscapes (so-called ``barren plateaus''),  
means that only a vanishingly small region of parameter space exists which has gradients large enough 
to measure with high enough precision to perform gradient descent.  This region of concentrated cost has been termed a ``narrow gorge.'' \cite{arrasmith_equivalence_2021} 
As a result, initializing the optimization from a random point in parameter space 
is bound to land in a barren plateau, 
meaning that the number of circuit executions (shots) needed to resolve the search direction 
increases exponentially with the number of qubits, 
preventing any opportunity for quantum advantage. 
While intelligent heuristics for parameter initializations might help
protect an optimization from getting stuck in a barren plateau 
(e.g., starting from a Hartree-Fock solution in molecular VQEs),
the success is largely determined on a case-by-case basis.\cite{mcclean_barren_2018}

In this work, we show that our recently introduced adaptive variational algorithm, ADAPT-VQE,\cite{grimsley_adaptive_2019} is largely immune to local minima and barren plateaus in the parameter landscape. Both issues are avoided because the algorithm systematically ``burrows'' a deep well in the landscape until the global minimum is reached. In other words, ADAPT-VQE dynamically modifies its parameter landscape in such a way that problematic regions are never explored. This phenomenon can be understood directly from the gradient criterion used to iteratively update the wavefunction ansatz. We illustrate this behavior with simulations of several different molecules. In Appendix \ref{adaptN}, we also show that the smoothness of the landscape can be controlled by intentionally overparameterizing the ansatz. 

\section{ADAPT-VQE Algorithm and Numerical Simulations}
\subsection{ADAPT-VQE}    
In recent work, we developed a dynamic framework for constructing ans{\"a}tze that have much faster energy convergence with respect to circuit depth.
This approach, referred to as ADAPT-VQE,\cite{grimsley_adaptive_2019,tang_qubit-adapt-vqe_2021}
 uses measurements of the molecular energy gradient to dynamically grow an ansatz, operator by operator, creating a highly compact ansatz that quickly converges to the exact solution.
Defining a pool of anti-Hermitian operators, $\mathcal{A} = \{A_i\}$, we outline the steps in Algorithm \ref{alg:adapt}.
\begin{algorithm}\label{alg:adapt}
\caption{ADAPT-VQE Algorithm}
$k\gets 0$\;
$\mathcal{U}_0\left(\boldsymbol\theta\right) \gets \mathbf{1}$\;
\While{Converged = False}{
    \For{$A_i \in \mathcal{A}$}{
       $\mathbf{g}_i \gets \left|\frac{\partial}{\partial \theta_i}\bra{0}\mathcal{U}_k^\dagger\left(\boldsymbol\theta\right)e^{-\theta_iA_i}\mathcal{H}e^{\theta_iA_i}\mathcal{U}_k\left(\boldsymbol\theta\right)\ket{0}\right|_{\theta_i = 0, \boldsymbol{\theta}} $\;
    }
    $i\gets$ index of largest element of $\mathbf{g}$\;
    $k\gets k+1$\;
    $\mathcal{U}_k\left(\boldsymbol\theta\right)\gets e^{\theta_i A_i}\mathcal{U}_k\left(\boldsymbol\theta\right)$\;
    $\boldsymbol{\mathbf{\theta}} \gets \underset{\boldsymbol\theta}{argmin} \bra{0}\mathcal{U}_k^\dagger\left(\boldsymbol\theta\right)\mathcal{H}\mathcal{U}_k\left(\boldsymbol\theta\right)\ket{0} $\;
    $E_k \gets \bra{0}\mathcal{U}_k^\dagger\left(\boldsymbol\theta\right)\mathcal{H}\mathcal{U}_k\left(\boldsymbol\theta\right)\ket{0} $\;
}
Return $E_k$, $\boldsymbol\theta$\;
\end{algorithm}

At each ADAPT-VQE iteration, the gradient, $\frac{\partial E}{\partial \theta_i}$, is measured with respect to all operators in the pool.
The operator with the largest gradient magnitude is then added to the ansatz with the associated parameter initialized to zero.  
The other parameters in the ansatz are initialized using the optimal values from the previous step (we refer to this as parameter ``recycling'').  
At this point, an ordinary VQE is performed using some classical optimization algorithm. 
In this work, we exclusively use the Broyden–Fletcher–Goldfarb–Shanno (BFGS) method\cite{fletcher_practical_2000}, a quasi-Newton strategy, because we are explicitly seeking information about local minima, and because we are not including any noise models in our simulations.  
Because we initialize the new parameter added during each ADAPT-VQE iteration to zero, the new trial circuit is equivalent to the previous one during the first VQE iteration.  Consequently, the energy can only improve during this VQE, i.e., the energy decreases monotonically.  Parameters are added one-by-one in this fashion until some convergence criteria are achieved.  Reasonable choices include the norm (either $l^2$ or $l^\infty$) of the vector of gradients, $\mathbf{g}$, or the number of operators in the ansatz.

All simulations were conducted using a locally developed code  which can be found on GitHub at \url{https://github.com/hrgrimsl/adapt}. OpenFermion\cite{mcclean_openfermion_2020} was used to construct matrix representations of operators under the Jordan-Wigner transformation and PySCF\cite{sun_pyscf_2018} was used to obtain molecular integrals.  
Because our focus in this work is to first understand the noise-free parameter landscapes associated with ADAPT-VQE, all simulations are performed without any noise models.  Future work will explore how the presence of noise affects the landscapes.
For all the ADAPT-VQE calculations in this work, the unitary coupled cluster with singles and doubles (UCCSD) operator pool is used,\cite{grimsley_adaptive_2019} 
without spin-complemented or spin-adapted operators.  
Details of this pool are provided in Appendix \ref{pool}.  

\subsection{Prevalence and Distribution of Local Minima}
In this section, we numerically explore the parameter landscapes of several example systems using ADAPT-VQE. 
Our aim is to characterize the way in which the number and distribution of local minima change as ADAPT-VQE gradually increases the length of the ansatz (and thus the depth of the circuit). 
For each molecule and bond distance considered, we first run ADAPT-VQE normally, 
where the initial parameter values used in the VQE at each iteration of the algorithm are chosen to be the ``recycled'' parameters, i.e., the optimal values obtained from the previous iteration. 
This yields an ansatz that reproduces the target ground state with high accuracy.

After using ADAPT-VQE to define the ansatz, 
we then use this ansatz to search for local minima by repeatedly reinitializing each VQE with randomly chosen parameters, and reoptimizing.\footnote{Each parameter was randomly initialized on the $[0,2\pi)$ interval in order to coincide with the period of $e^{A_i\theta_i}$ for the chosen pool.}  
In this work, we performed 300 such random initializations for each ansatz considered. 
For each layer of the ansatz and each random initialization, we record the minimum energy obtained by the VQE subroutine. These values correspond to the energies of local minima in the landscape associated with each ansatz.

In addition to these random initializations, we also include both the ``recycled'' parameters from the previous VQE (the default initialization in ADAPT-VQE\cite{grimsley_adaptive_2019}) and the $\mathbf{0}$ parameter vector associated with the Hartree-Fock (HF) reference.  
All 302 initializations of a given ansatz are then optimized with BFGS, and the resulting energy errors are shown with rainbow-colored bars in each figure.  
The colors indicate relative energy ordering at a given ansatz, 
such that red corresponds to the highest energy and violet to the lowest energy.  The recycled initialization's outcome is of particular interest since this is the default, 
deterministic initialization for ADAPT-VQE, 
and the approach used when growing the ans{\"a}tze used in the data. 
These conventions will be used throughout this work.

We consider linear H$_4$ and H$_6$ at 1 and 3 \AA{} as toy models exhibiting varying degrees of electron correlation (and entanglement in the target wavefunction).  In addition, we study LiH at 1.62 \AA{} and BeH$_2$ at 1.33 \AA{} as examples of real molecules at equilibrium geometries.  These geometries were obtained through optimization at the B3LYP \cite{becke_densityfunctional_1993}/6-31G$^*$\cite{dill_selfconsistent_1975, ditchfield_selfconsistent_1971, hariharan_influence_1973, hehre_selfconsistent_1972} level of theory in PySCF,\cite{sun_pyscf_2018} and are given in the Supplementary Information. 
All ADAPT-VQE calculations were performed in the STO-3G\cite{hehre_selfconsistent_1969,collins_selfconsistent_1976} basis.  
\begin{figure}
\includegraphics[width=.48\textwidth]{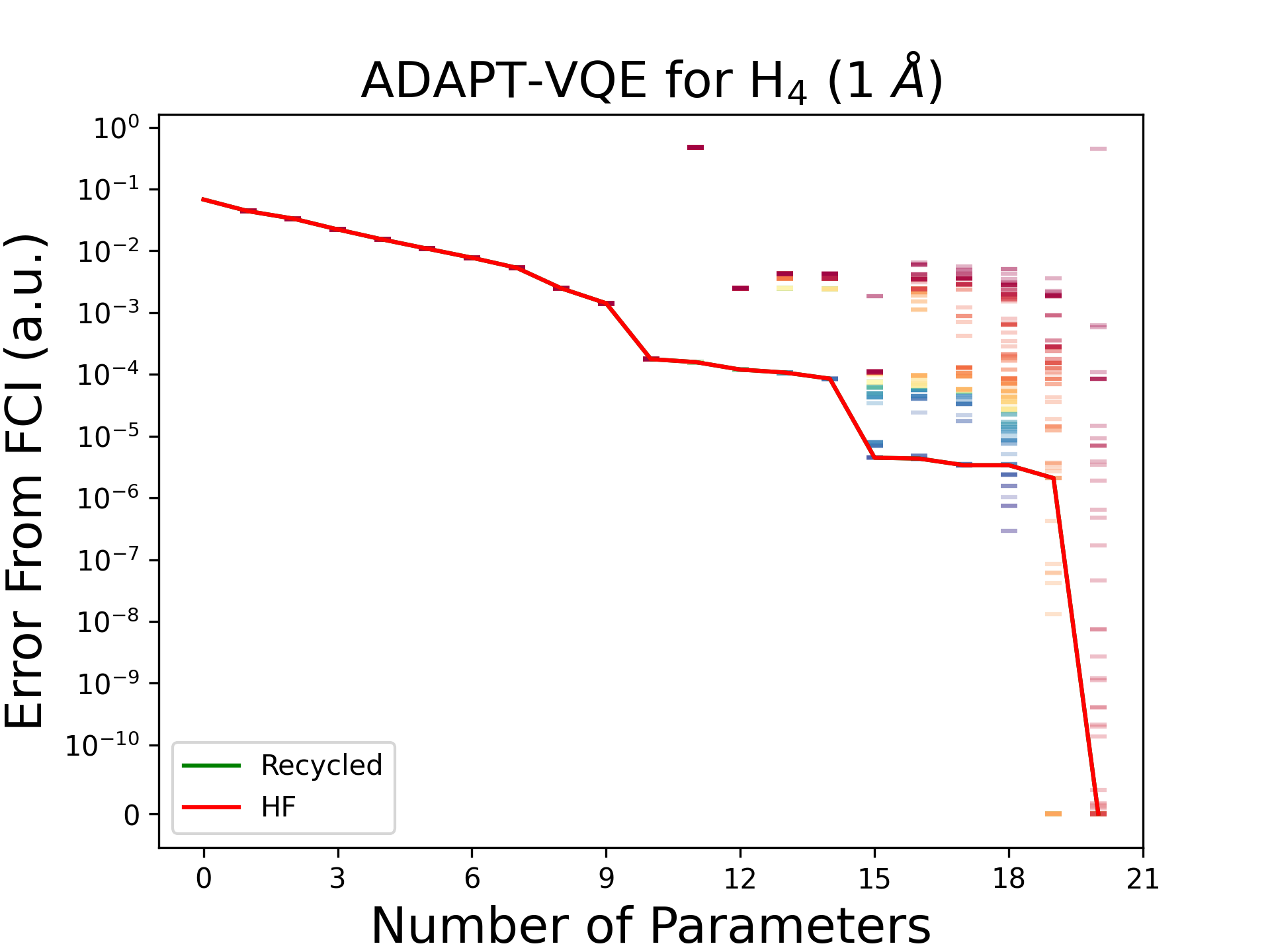}
\caption{ADAPT-VQE Results for H$_4$ at 1 \AA.  The $x$-axis corresponds to the number of ADAPT-VQE iterations, i.e. the number of operators in the ansatz at a given step.  The $y$-axis corresponds to the error from the exact FCI energy.  The red curve corresponds to the energy obtained through BFGS minimization using an HF guess, i.e. one where all parameters are zero.  The green curve corresponds to the energy obtained through BFGS minimization using the standard ADAPT-VQE in which optimal parameter values in one iteration are recycled as initial guesses in the next iteration, and with the new parameter initialized to zero.  The colored dots correspond to all the energies obtained through BFGS optimizations, with red being the highest energy and violet the lowest.}\label{h4_1A_adapt}  
\end{figure}
\begin{figure}
\includegraphics[width=.48\textwidth]{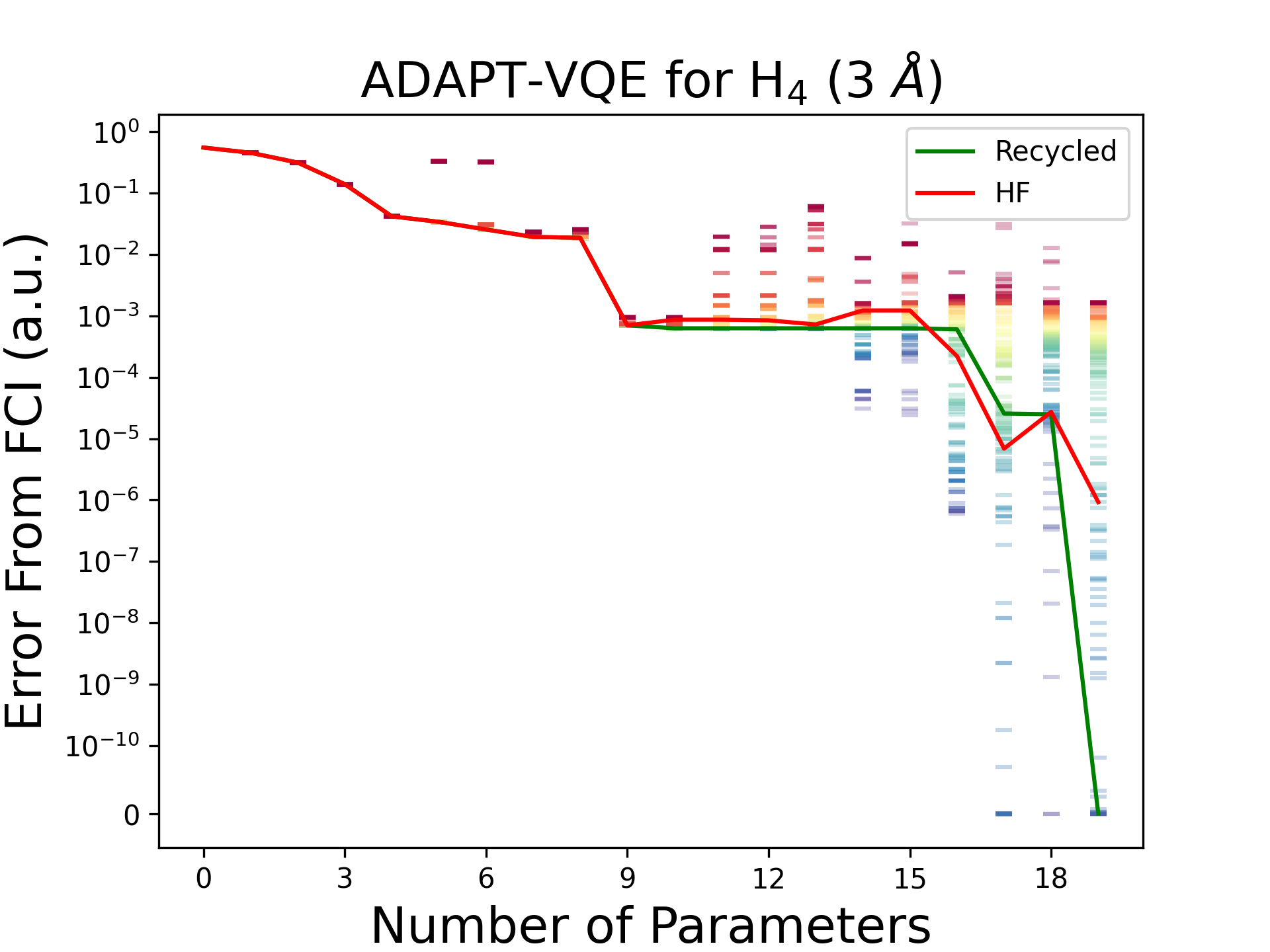}
\caption{ADAPT-VQE Results for H$_4$ at 3 \AA.  The axes and colors are as in Fig. \ref{h4_1A_adapt}.}\label{h4_3A_adapt}  
\end{figure}

\paragraph*{$\text{H}_4$ molecule} In Fig. \ref{h4_1A_adapt} we show the energies (relative to the global minimum obtained from a full configuration interaction (FCI) calculation) of the various local minima as a function of ansatz length (as defined by the ADAPT-VQE algorithm). 
After a short period without local minima, 
the random initializations begin to converge on an increasing number of distinct local minima as the number of parameters increases. 
In contrast to the random initialization, both the HF and the recycled initializations converge to the same  minimum for H$_4$ at 1 \AA{}, which is consistently  better than the average random initialization.  
This is our first indication that good initializations can reliably avoid high-energy traps.  Interestingly, even though ADAPT-VQE doesn't always find the lowest energy trap, it does eventually converge. 
Additionally, we observe that there are still many local minima even after these ``chemically informed'' guesses are able to reach the exact ground state.
In Appendix \ref{adaptN}, we consider the prospect of removing local minima through systematic overparameterization for H$_4$ at 1 \AA{}.  While we are successful in removing local minima using a novel ``ADAPT$^N$'' approach, deeper circuits are actually required to achieve the overparameterization than to simply add operators until ADAPT-VQE reaches the ground state in spite of local minima.  

In Fig. \ref{h4_3A_adapt}, we see that for the more strongly correlated 3 \AA{} bond distance, the HF and recycled initializations differ.  The recycled initialization is able to reach the ground state with fewer parameters than the HF initialization, suggesting that it is generally a more robust choice than HF for finding the global minimum.  
Again, we see ADAPT-VQE converging to the exact solution far faster than a typical (yellow-green) random initialization. 

\begin{figure}
\includegraphics[width=.48\textwidth]{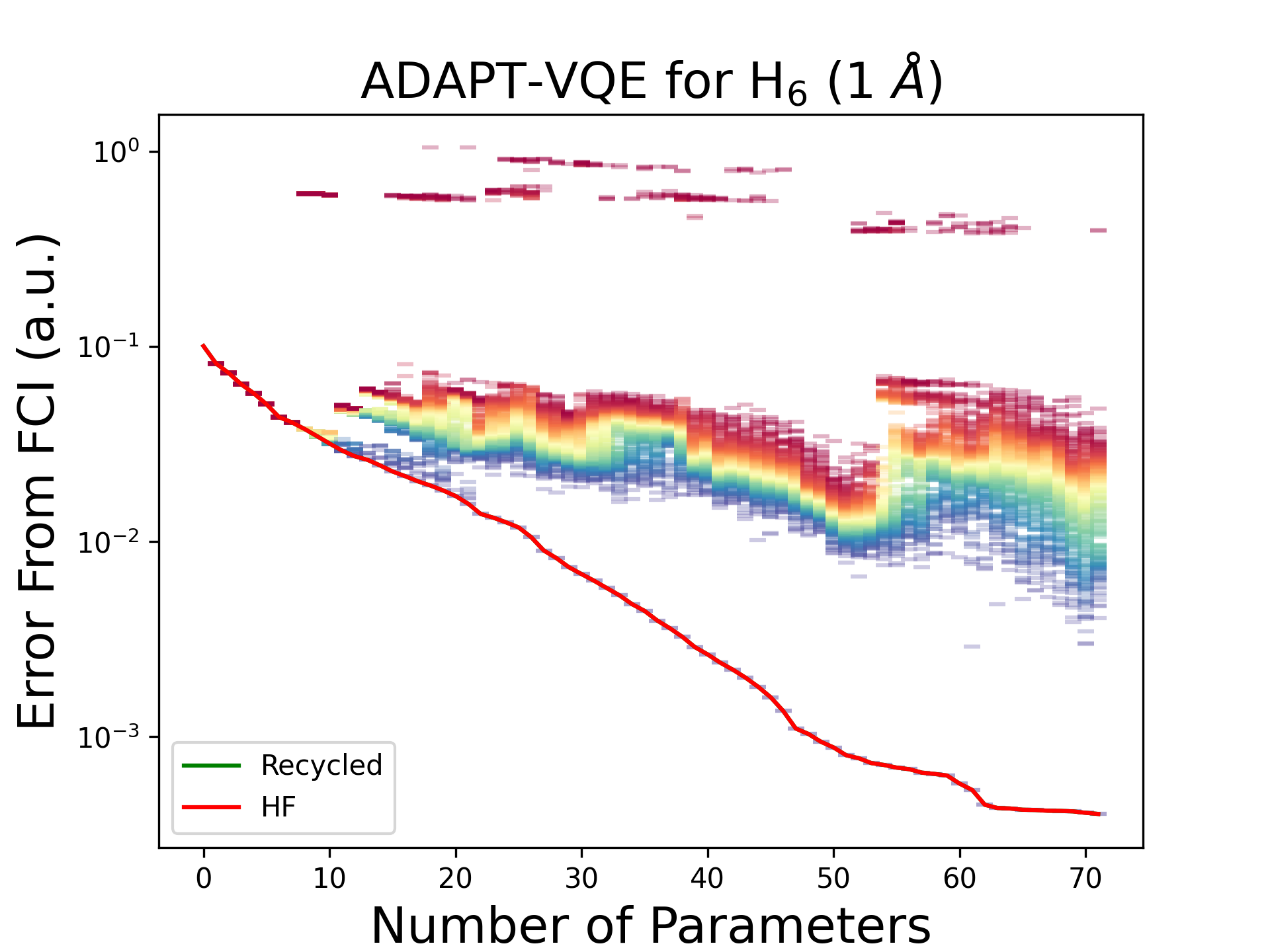}
\caption{ADAPT-VQE Results for H$_6$ at 1 \AA.  The axes and colors are as in Fig. \ref{h4_1A_adapt}.}\label{h6_1A_adapt}  
\end{figure}
\begin{figure}
\includegraphics[width=.48\textwidth]{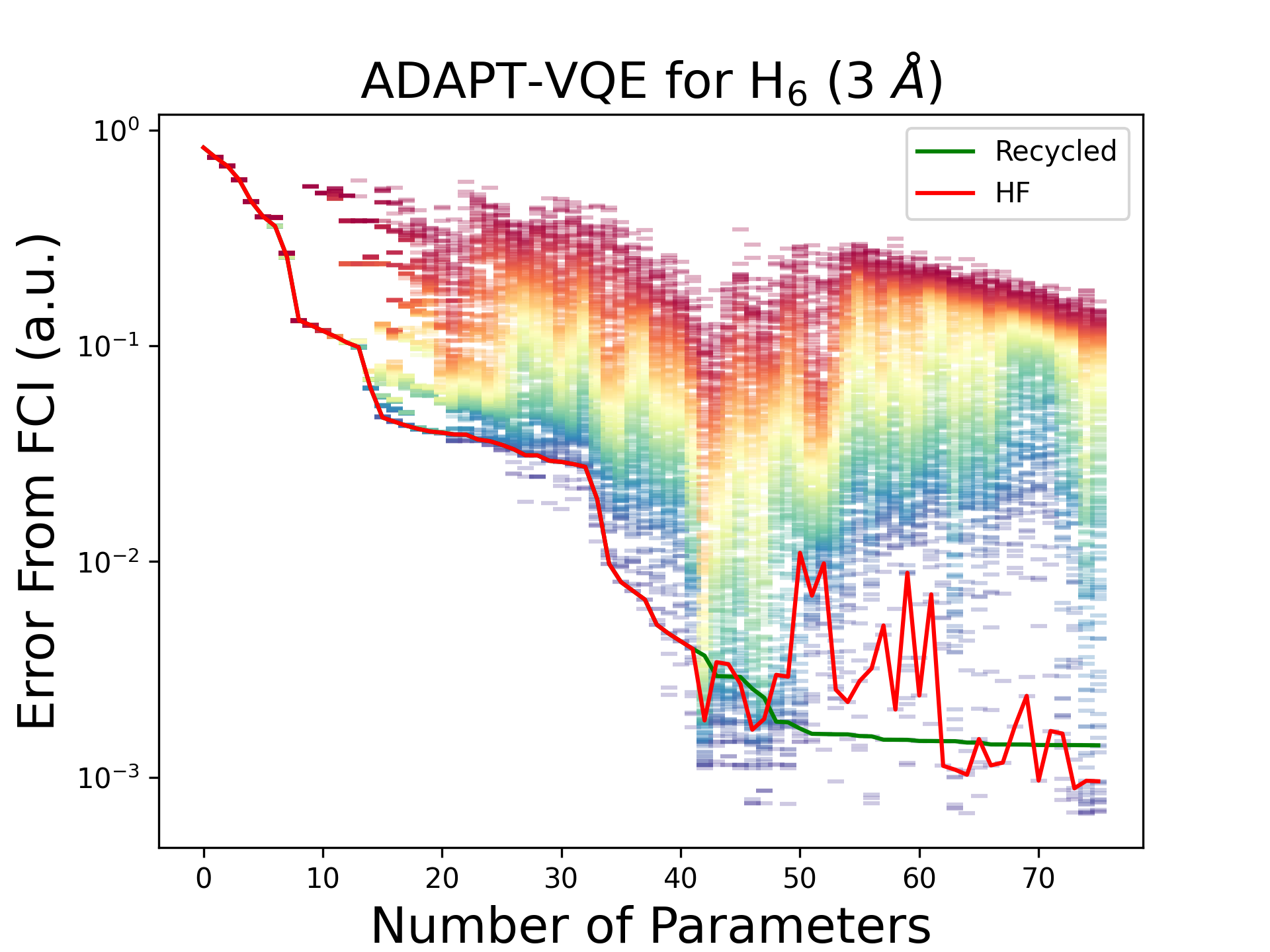}
\caption{ADAPT-VQE Results for H$_6$ at 3 \AA.  The axes and colors are as in Fig. \ref{h4_1A_adapt}.}\label{h6_3A_adapt}  
\end{figure}

\paragraph*{$\text{H}_6$ molecule } In Fig. \ref{h6_1A_adapt}, we begin to see the true power of an intelligent guess by simulating H$_6$ at 1\AA{}.  As the ansatz grows longer, a  massive gap opens up between the random guesses and the HF/recycled ones.  This gap implies that in practice, it is very difficult to do better than simply recycling the previous parameters in ADAPT.  
This gap is further numerical evidence of a ``narrow gorge'', in which the exact solution is hypothesized to exist.\cite{arrasmith_equivalence_2021} Although such a landscape is often associated with optimization difficulties, here we see that ADAPT-VQE is able to stay very close to the narrow gorge, avoiding such issues.
We emphasize that this feature is not only a result of good initialization~\cite{skolik_layerwise_2021}, but rather a cooperative effect between initialization and the gradient-guided ansatz construction. 
In Appendix \ref{sophia}, we demonstrate this explicitly by performing simulations using the recycled initialization, but on randomized (not gradient-guided) ans\"atze.

In Fig. \ref{h6_3A_adapt}, the energy distribution of the local traps significantly increases. 
While we no longer see such an obvious gap for H$_6$ at 3 \AA,  the HF and recycled initializations are still far better than random ones. 
As the ansatz grows in depth (i.e., around 45 parameters), 
we notice a quick rise in the median energy of the traps found. 
This indicates that as the number of parameters increases, so too does the number of local traps. 
Furthermore, these new traps are preferentially high in energy, thus moving the median solution to higher energies. 
This further implies that as the system grows in size, the overwhelming number of solutions will be high in energy, making random sampling of VQE initializations intractable. 
Focusing on the comparison between HF and the recycled initialization, we see that they begin to differ significantly (around 50-60 parameters) where the recycled guess outperforms the zero initialization (HF).  
The improvement over HF is unsurprising given the strong correlation in this system, which ADAPT-VQE can handle.  

\begin{figure}[hbtp!]
\includegraphics[width=.48\textwidth]{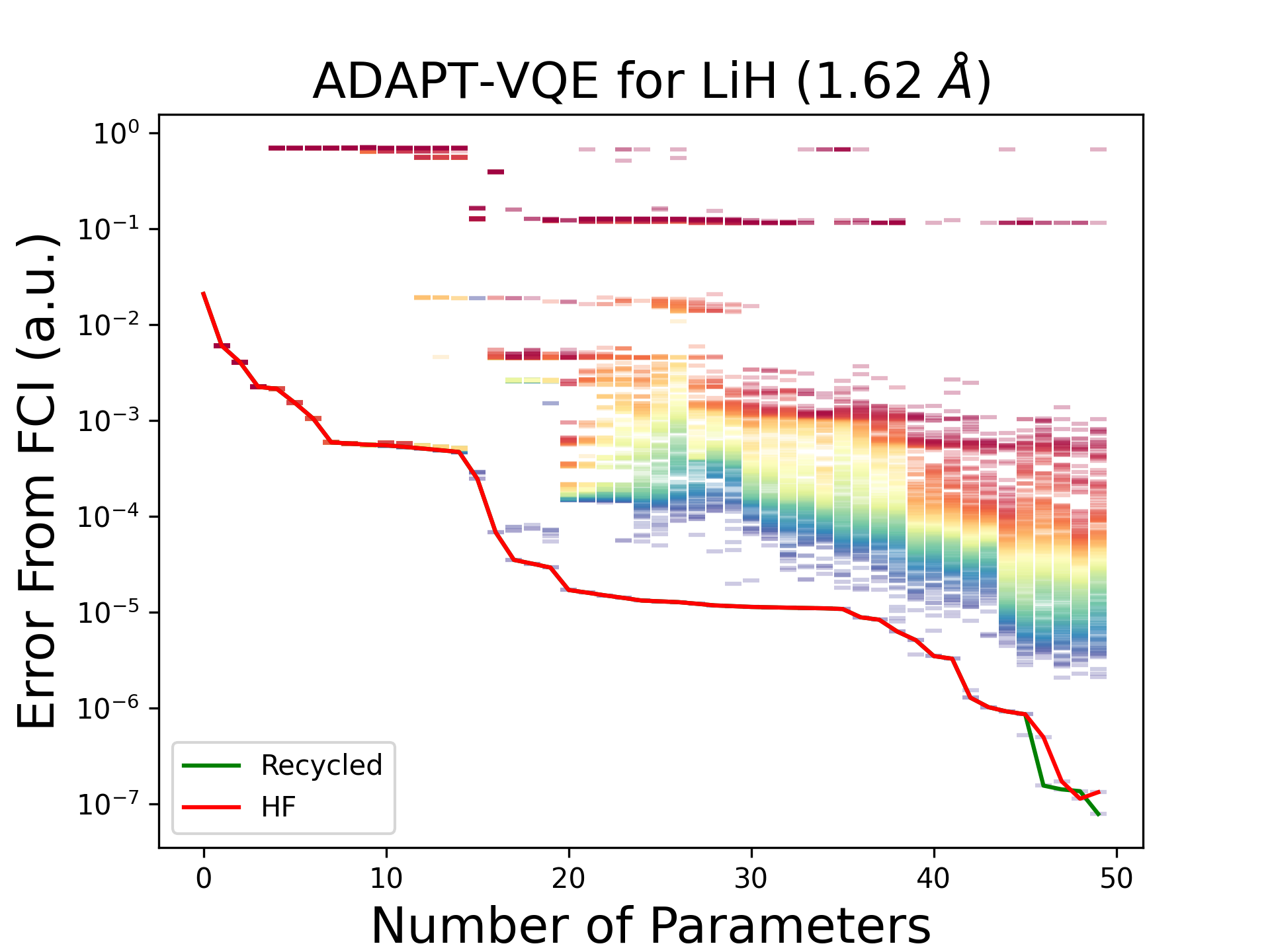}
\caption{ADAPT-VQE Results for LiH at 1.62 \AA.  The axes and colors are as in Fig. \ref{h4_1A_adapt}.}\label{lih_adapt}
\end{figure}

\paragraph*{$\text{LiH}$ molecule } In Fig. \ref{lih_adapt} we see similar behavior for LiH to that of H$_6$ at 1 \AA{}.  While the solution gap is less pronounced, both HF and the recycled initialization are always significantly better than nearly every random initialization.  

\begin{figure}[hbtp!]
\includegraphics[width=.48\textwidth]{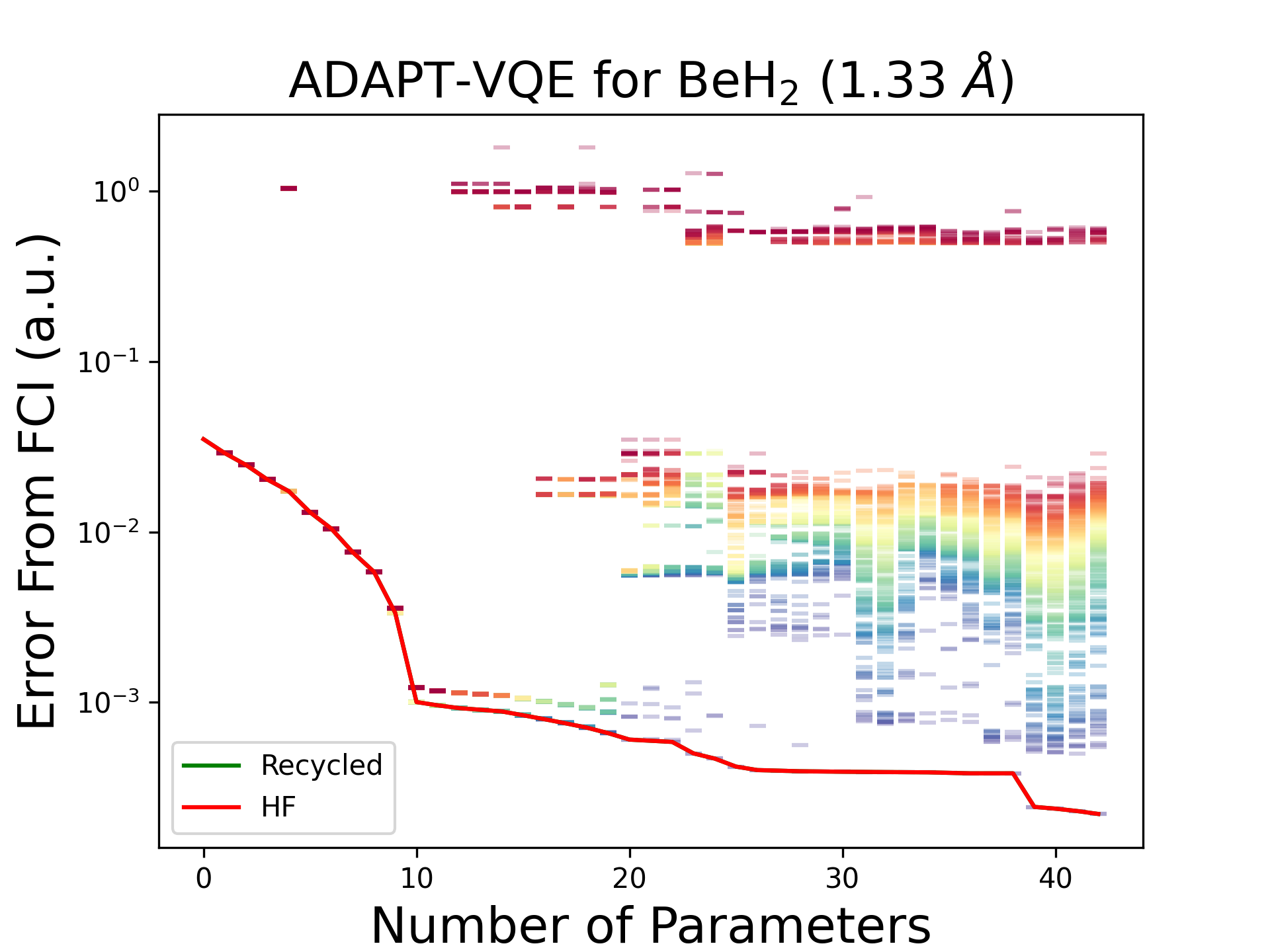}
\caption{ADAPT-VQE Results for BeH$_2$ at 1.33\AA.  The axes and colors are as in Fig. \ref{h4_1A_adapt}.}\label{beh2_adapt}  
\end{figure}
\paragraph*{$\text{BeH}_2$ molecule } We observe similar behavior once again in Fig. \ref{beh2_adapt} for BeH$_2$.  In this relatively weakly correlated system, the HF and recycled curves overlap completely, and once again outperform the random initialization strategy.

In all cases, we observe that for more than a few parameters, local minima emerge, and for large numbers of parameters, these minima often dominate the energy landscape.  In many cases initializing all parameters to 0 (HF) is a reasonable choice that leads to low energy minima. Unsurprisingly, however, the quality of HF initialization diminishes in strongly correlated systems like H$_4$ and H$_6$ at 3 \AA{}.  In such cases, we expect the optimal parameters to be further from 0 since the mean-field HF solution does not accurately represent the wavefunction.  For these two strongly correlated systems, the recycled ADAPT-VQE parameters are more reliable as initial guesses, and we expect the advantage to become more pronounced as system sizes increase for strongly correlated systems.

\begin{figure}
\includegraphics[width=.48\textwidth]{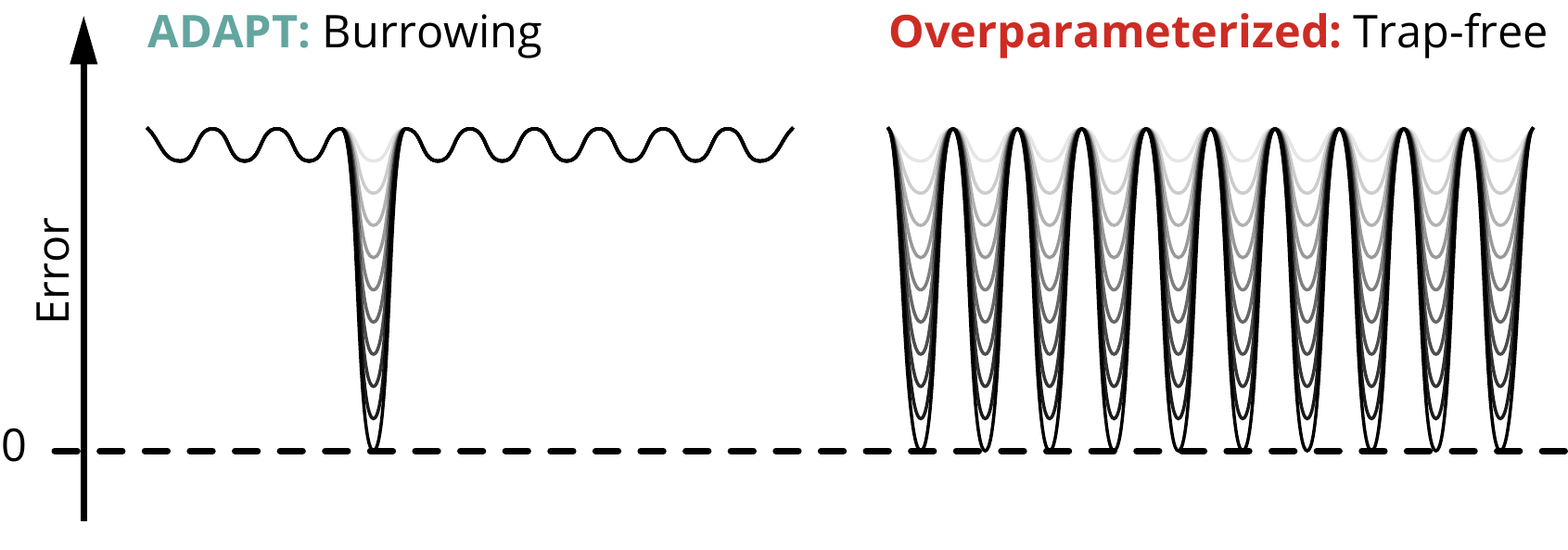}
\caption{Schematic illustration of how the parameter landscapes change as parameters are added using a) ADAPT-VQE,
and b) a ``controllable'' or overparameterized ansatz which has a greater number of parameters than the rank of the DLA. }\label{fig:burrow} 
\end{figure}

\paragraph*{Trap ``Burrowing''} The problem of local minima seems to be partially mitigated by ADAPT-VQE itself.  
Even in cases where the recycled initialization converges to a high-energy trap, ADAPT-VQE progresses by adding an operator which is chosen to preferentially deepen the current trap (via the gradient criterion). As such, over a sequence of ADAPT-VQE iterations, the current trap becomes increasingly deep relative to the other parameter traps, such that a gap can open up between the current minimum (which approaches the global minimum) and all other local minima.  Thus ADAPT-VQE appears to ``burrow'' into the parameter landscape, creating a single deep well as opposed to stabilizing all local minima (i.e., reaching overparameterization). 
This burrowing effect is depicted graphically in Fig. \ref{fig:burrow}.

\subsection{Sensitivity to Parameter Landscape Gradients}
\subsubsection{Insensitivity to Barren Plateaus}
In the previous section, we demonstrated that while the parameter landscapes exhibit a large number of local traps that are high in energy, ADAPT-VQE is robust due to the fact that any local minimum in early stages of the algorithm can be deepened into a global minimum at later stages. 
This same mechanism implies a similar robustness to the presence of barren plateaus.
As mentioned above, the barren plateau phenomenon has been recently recognized as a serious obstacle to the use of VQEs in practical settings. 
The problem arises from the observation that highly expressive ans{\"a}tze (more specifically, circuits which form a 2-design), which are attractive from an accuracy perspective, exhibit an exponentially decreasing gradient variance
with increasing system size. 
This means that the vast majority of parameter space becomes essentially flat.
In the course of optimizing the parameters of such an expressive ansatz, a randomly chosen initialization 
will (with overwhelming probability) correspond to a point in parameter space where the gradient of the cost function is so small that an exponentially large number of measurements are needed to resolve a meaningful search direction in the presence of noise.
As a result, the ability to optimize or train such expressive circuits is suspect at best. 
While a physically inspired parameter initialization can be effective (e.g., HF initialization), difficult cases (like those exhibiting strong correlation) may prevent efficient initialization. 

Unlike the non-adaptive situation in which a static ansatz is first defined and then optimized, ADAPT-VQE slowly brings a given stationary point (initially the reference state) to the exact solution, via this burrowing mechanism. 
As such, each VQE performed along the way is ``warm-started'', in that one already has a decent initialization coming from the previous optimization. 
Using this recycled initialization, we have a clear characterization of the parameter landscape about the initial point: all previous parameters are optimized, and thus have zero gradients, and the newly added operator has a large gradient by design, since we specifically add the operator with the largest gradient. 
This means that each VQE subroutine in the ADAPT-VQE algorithm is initialized with a single parameter which is guaranteed to be greater than $\epsilon$ (the ADAPT-VQE convergence threshold). 
Based on this argument, we do not expect difficulty due to barren plateaus when training ADAPT-VQE ans\"atze as system sizes are scaled up. 
We emphasize that this argument does not suggest that the ans\"atze constructed by ADAPT-VQE are free from barren plateaus, 
only that our algorithm remains localized to a region in parameter space with significant gradients.

\begin{figure}[hbt!]
\includegraphics[width=\linewidth]{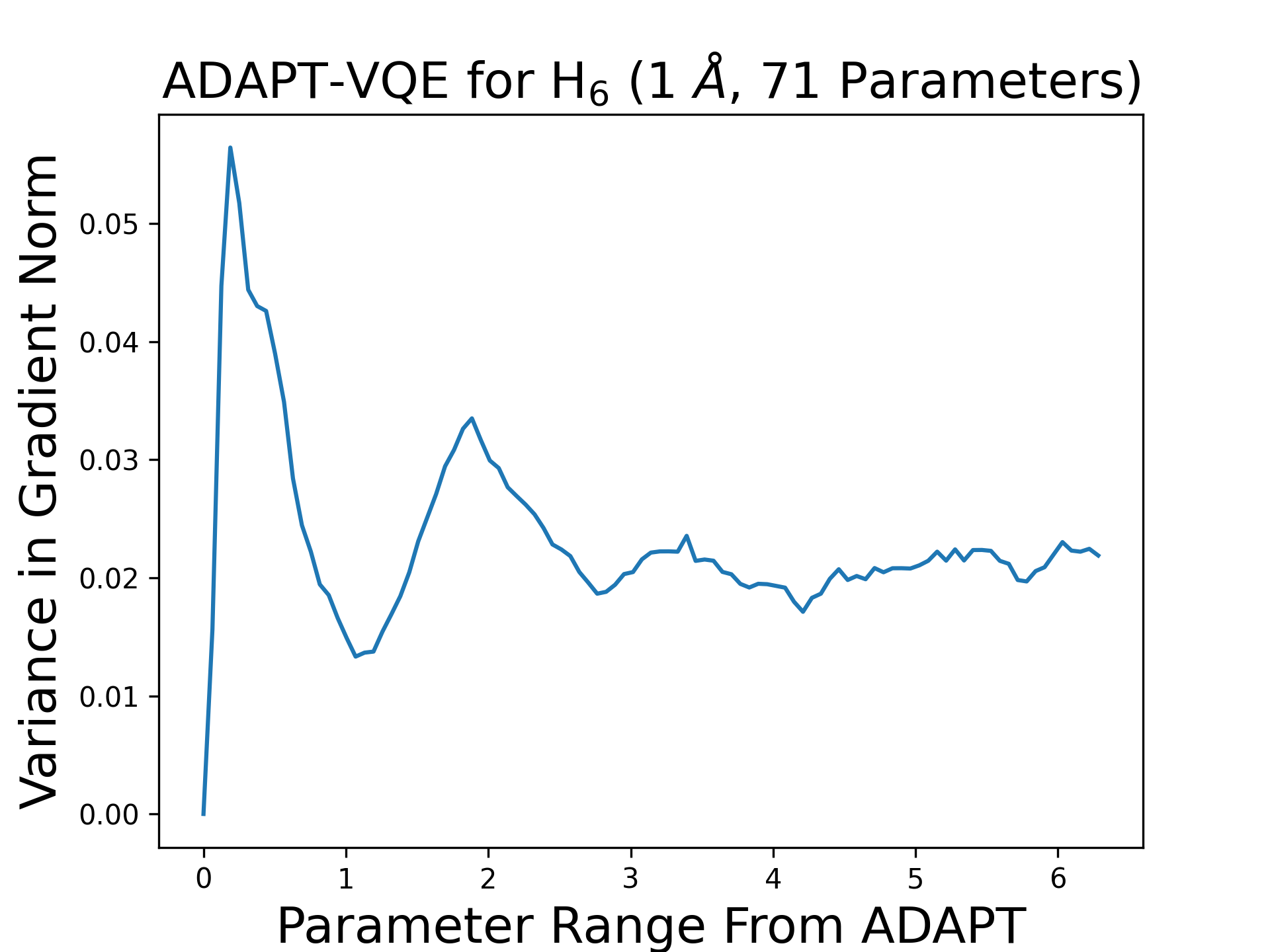}
\caption{Variance of the gradients for the 1 \AA{} H$_6$ ansatz at 71 parameters as a function of parameter search region.  Parameter vectors were selected from a 71-dimensional hypercube with each parameter randomly chosen within a certain adjustable range ($x$-axis), i.e. at $x=2\pi$, each parameter is chosen from the interval [$\theta^*-2\pi,\theta^*+2\pi$), where $\theta^*$ is the recycled ADAPT-VQE solution parameter.}\label{h6_1A_bp}
\end{figure}

In Fig. \ref{h6_1A_bp}, we show that the landscape is steeper near the minimum compared to points further away in parameter space.
Here, the variance of the gradient is obtained by sampling the parameter space within a hypercube of increasing volume, centered at the ADAPT-VQE default initialization. 
As the side of this hypercube increases (moving right on the $x$-axis), we plot the gradient variance of an increasing reach in parameter space. 
We find that near the default initialization defined by ADAPT-VQE, the landscape has large gradients. 
However, as the search space is increased such that we randomly consider arbitrary points in parameter space, 
the gradient variance flattens out.

\subsubsection{``Gradient Troughs''}\label{sec:gradtrough}
Although barren plateaus seem to pose no threat to the ability to scale up ADAPT-VQE based on the arguments in the previous section, 
there is still a related issue that might prevent ADAPT-VQE from converging to accurate solutions. 
As described above, at each ADAPT-VQE step, the ansatz is extended using the operator with the largest gradient:
\begin{align}
    \frac{\partial E}{\partial \theta_i} &= \bra{\psi(\mathbf{\theta})}[\hat{H},\hat{A}_i]\ket{\psi(\mathbf{\theta})}.
\end{align}
The ansatz is then repeatedly extended until the largest gradient in the operator pool\footnote{In the first paper the convergence criterion was taken to be the norm of the gradients in the pool, rather than the maximum.}
is smaller than some threshold, $\epsilon$.
Noise on a NISQ device, however, defines some lowest possible threshold, $\epsilon_{\text{min}}$, that can be resolved using a given shot allowance.
In our earlier work,\cite{grimsley_adaptive_2019} we sometimes observed non-monotonic convergence of the gradients as a function of ansatz length (although the energy convergence is guaranteed to be monotonic), 
such that as the ansatz is extended, the pool gradients might first decrease, then increase again before finally converging. 
This ``gradient trough'', therefore presents a challenge in the presence of noise. 
If a gradient trough appears and drops below the NISQ resolvable threshold, $\epsilon_{\text{min}}$,
then the ADAPT-VQE algorithm may halt prematurely.

\begin{figure}
\includegraphics[width=\linewidth]{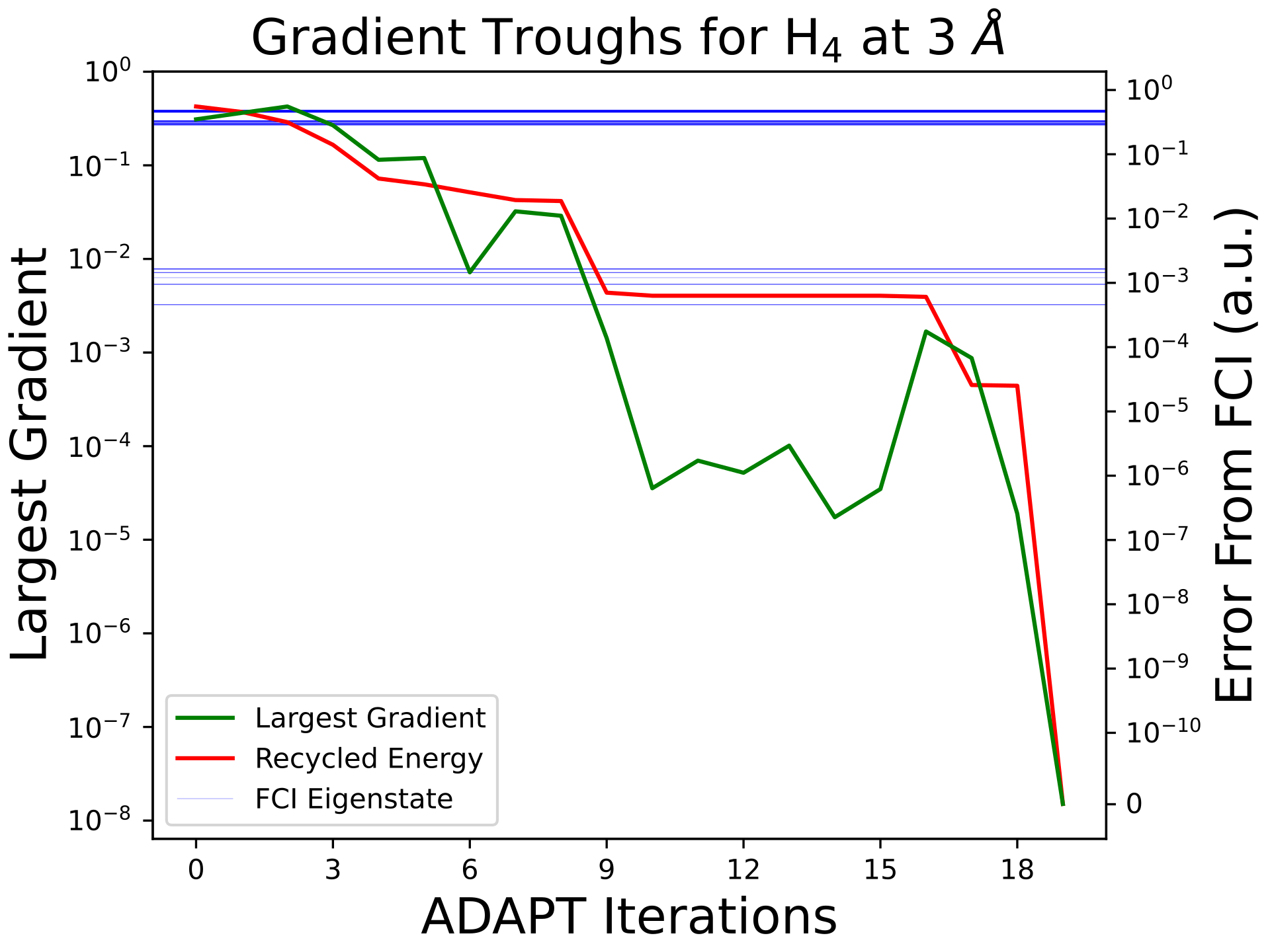}\caption{ADAPT-VQE with recycled parameters for H$_4$ at 3 \AA{}.  The $x$-axis corresponds to the ADAPT-VQE iteration.  The green curve depicts the gradient associated with the operator to be added at each step, while the red curve depicts the energy at each ADAPT-VQE step.  The blue lines depict the excited FCI eigenstates which are lower than the HF energy.}
 \label{h4_3A_grad_trough}
\end{figure}

\begin{figure*}
\includegraphics[width=\textwidth]{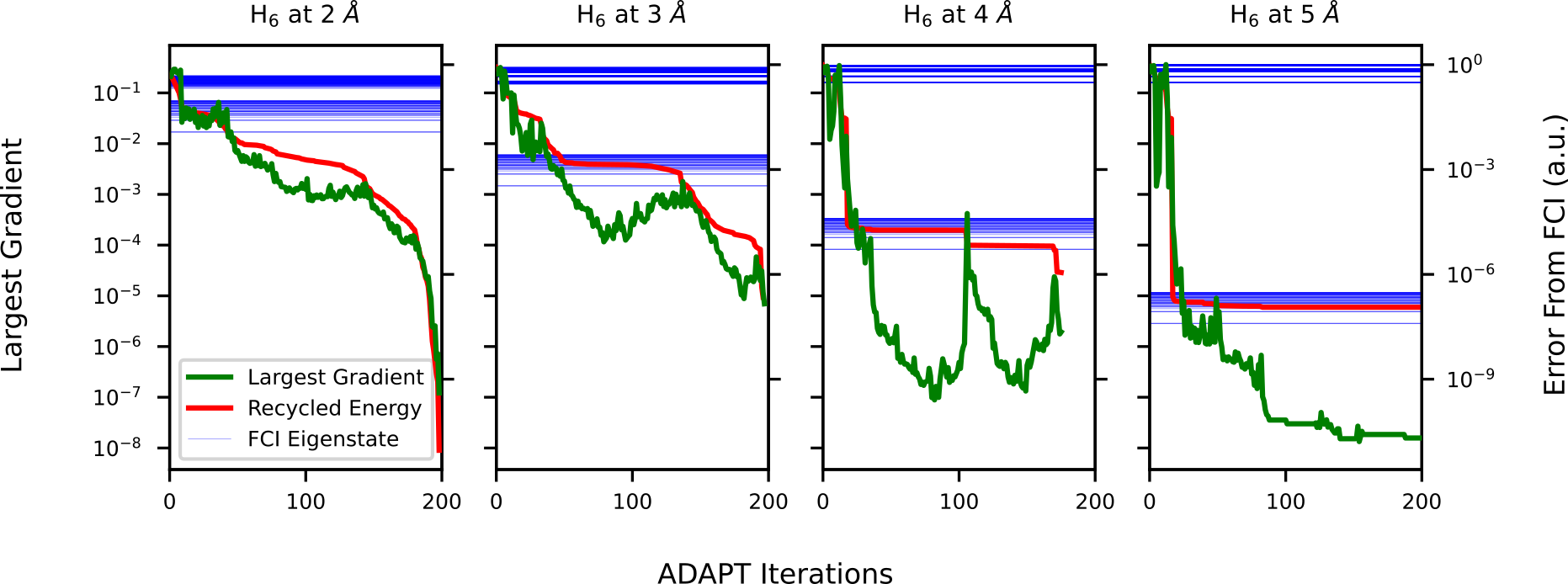}
\caption{ADAPT-VQE with recycled parameters for H$_6$ at 3 \AA{}.  The $x$-axis corresponds to the ADAPT-VQE iteration.  The green curve depicts the gradient associated with the operator to be added at each step, while the red curve depicts the energy at each ADAPT-VQE step.  The blue lines depict the excited FCI eigenstates which are lower than the HF energy.}\label{all_troughs}
\end{figure*}

How do these gradient troughs grow with system size?
If we were to find that they grow exponentially fast, 
meaning that the largest gradient in the operator pool is exponentially suppressed as the number of qubits increases, then this would suggest concern for the scalability of ADAPT-VQE. 
However, this does not need to be the case.
Choosing a local orbital basis\footnote{One is always free to rotate occupied or virtual orbitals 
without changing the associated Slater determinant, due to orbital subspace rotational invariance.}
one can imagine trivial situations where the gradients not only avoid exponential suppression, but any suppression at all.
Consider the $n^{th}$ iteration of an ADAPT-VQE calculation of a molecular wavefunction, $\ket{\psi_n}$.
If one were to double the number of qubits by adding another molecule (at infinite distance
so as to remove interactions between the systems), 
the total wavefunction at iteration $2n$ would  have a product form, $\ket{\psi^{AB}_{2n}}=\ket{\psi^A_n}\ket{\psi^B_n}$.
Any pool operator $\hat{O}_i$ that is local to either subsystem has the exact same gradient 
in the supersystem, $\ket{\psi^{AB}_{2n}}$, as it does in the subsystem, $\ket{\psi^{A}_{n}}$.
For example, consider an operator, $\hat{O}_i^A$, local to subsystem $A$:
\begin{align}\label{eq:gradient_scaling}
    \frac{\partial E^{AB}}{\partial \theta_i}  =& \mel{\psi^{AB}_{2n}}{[\hat{H},\hat{O}^A_i]}{\psi^{AB}_{2n}} \nonumber\\
     =& \mel{\psi^{AB}_{2n}}{[\hat{H}^A + \hat{H}^B ,\hat{O}^A_i]}{\psi^{AB}_{2n}} \nonumber\\
     =& \mel{\psi^{A}_{n}}{[\hat{H}^A ,\hat{O}^A_i]}{\psi^{A}_{n}} \ip{\psi^{B}_{n}}{\psi^{B}_{n}} \nonumber\\
     =& \frac{\partial E^{A}}{\partial \theta_i}.
\end{align}
The additive separability of non-interacting subsystems is referred to as ``size-consistency'' 
in the chemistry literature.
However, in addition to additive separability of the energy, 
size-extensive wavefunctions (like UCCSD) also demonstrate ``size-intensivity'' for intensive properties
(e.g., density, optical gaps, etc). 
As shown in Eq. \ref{eq:gradient_scaling}, the gradient with respect to a local rotation
is not affected by the presence of an additional non-interacting system, thus demonstrating size-intensivity.
In the limit of a large system, any further additions to the system size will necessarily be too far away from
a given subsystem to interact. 
Based on this argument, we don't expect gradient troughs to deepen asymptotically with system size.
However, more work is needed to characterize the behavior of gradient troughs as the system size increases in the presence of interactions.


\subsubsection{Effect of Low-Lying FCI Eigenstates}\label{spectral}
In order to understand the nature of the ``gradient troughs'' 
discussed in Sec. \ref{sec:gradtrough}, and shown in Figures \ref{h4_3A_grad_trough} and \ref{all_troughs}, we superimposed the low-lying FCI energies with the ADAPT-VQE energies computed.
The FCI spectrum is plotted as a set of blue horizontal lines. 
We only plot H$_4$ and H$_6$, as the other systems studied have no nearby excited states, nor do they exhibit any gradient troughs. 
In the region of the gradient trough,
the energy also becomes very flat,
(i.e., consider operators 9-16 in Fig. \ref{h4_3A_grad_trough} and 
operators 50-100 in Fig. \ref{all_troughs}).

By plotting the exact eigenstates on top of these curves, 
one readily sees that the gradient troughs occur when ADAPT-VQE falls inside of a nearly degenerate manifold of FCI excited states.
Should the ADAPT-VQE threshold be chosen loose enough (or if there is too much device noise to measure the gradient below this value) that the algorithm is aborted in this region, 
then ADAPT-VQE will be unable to advance further toward the ground state, 
remaining stuck as an approximation to an excited state (or in general some arbitrary superposition of the nearly degenerate eigenstates). 
This appearance of gradient troughs was first noticed 
in the paper that introduced ADAPT-VQE,\cite{grimsley_adaptive_2019} however the origin of the onset and the interpretation was not clear at that time.

As a consequence, although ADAPT-VQE isn't expected to suffer from the more general problem of barren plateaus,
more work is needed to understand how to escape any gradient troughs to ensure smooth convergence to the exact solution,
particularly when noise is included. 
This remains an outstanding problem associated with ADAPT-VQE,
warranting more research.

\section{Conclusions}
Underparameterized ans{\"a}tze are difficult to optimize due to large numbers of local minima, while highly expressive ans{\"a}tze are difficult to optimize due to barren plateaus. 
In this paper, we find that ADAPT-VQE does not suffer from these challenges. We have studied the parameter landscapes arising from various ADAPT-VQE generated ans\"atze and have arrived at the following conclusions:
\begin{enumerate}
\item \textbf{Recycled initialization helps avoid traps}: ADAPT-VQE's process of re-using parameters at each step focuses the search space on a local region, keeping the algorithm relatively easy to train despite the rough overall landscape. The parameter vector from the previous iteration tends to be a relatively good initial guess for the following ADAPT-VQE iteration. This means that by simply ``recycling'' the parameters from one ADAPT-VQE iteration to the next, the vast majority of parameter traps are entirely avoided.
\item \textbf{Trap burrowing corrects local minima}: Even if the early iterations get stuck in a trap, the adaptive construction iteratively extends the ansatz in a direction that is guaranteed to improve the cost function near the current stationary point.
By continually focusing on a local point in parameter space, ADAPT-VQE can ``burrow'' into a given local minimum, even if the vast number of traps remain high in energy.

\item \textbf{Barren plateau avoidance}: The nature of the ADAPT-VQE algorithm suggests that barren plateaus should not prove problematic in the parameter optimization step.  This originates from the fact that ADAPT-VQE specifically adds  a large gradient operator, generating a steep landscape, such that a search direction is resolvable without an exponential number of shots.
\item \textbf{Gradient troughs}: ADAPT-VQE can still exhibit numerical challenges. An exponentially vanishing \textit{pool} operator gradient could potentially arise, resulting in ADAPT-VQE becoming stuck during the operator addition step (in contrast to the parameter optimization step).
Numerical evidence suggests that these gradient troughs appear when the ADAPT-VQE energy starts to converge near one or more excited states. Heuristics for diagnosing and addressing such issues will be the focus of future work. 

\end{enumerate}

Despite the presence of local minima and the possibility of barren plateaus in standard ADAPT-VQE ansatze, we conclude that ADAPT-VQE can be optimized reasonably well through parameter recycling.
Consequently, in addition to being parameter- and gate- efficient,  ADAPT-VQE appears to be relatively immune to the problems of both local minima and barren plateaus in VQEs.

\section{Acknowledgements}
N.J.M., S.E.E., and E.B. are grateful for financial support provided by the U.S. Department of Energy. N.J.M. and E.B. acknowledge Award No. DE-SC0019199. 
S.E.E. acknowledges the DOE Office of Science, National Quantum Information Science Research Centers, Co-design Center for Quantum Advantage (C2QA), Contract No. DE-SC0012704. 
H.R.G. acknowledges support provided by the Institute for Critical Technology and Applied Science at Virginia Tech.
The authors thank the Advanced Research Computing at Virginia Tech for the computational infrastructure.

\appendix
\section{Operator Pool}\label{pool}
In this work, all experiments were performed with a pool of particle-hole (i.e., not generalized) Fermionic operators.  These operators were restricted to single excitations of the form $\hat{a}_i^a - \hat{a}_a^i$ and double excitations of the form $\hat{a}_{ij}^{ab}-\hat{a}_{ab}^{ij}$, where $i$ and $j$ are occupied spin orbitals and $a$ and $b$ are virtual spin orbitals.  While these operators are not required to respect $\hat{S}^2$ symmetry, the pool does not include ``spin flips,'' operators which change the total numbers of alpha and beta electrons.  This pool grows quartically with system size.  If these operators are used as generators for an ansatz of the form $\ket{\Psi} = e^{\theta_1 A_1}e^{\theta_2 A_2}\dots e^{\theta_M A_M}\ket{\phi_0}$ with infinite repetition of the operators and correct ordering, the FCI solution should be reachable.\cite{evangelista_exact_2019}  
However, we have not yet proven that the ADAPT-VQE convergence criterion goes to zero \textit{only} when converged to an exact eigenstate.

\section{\texorpdfstring{ADAPT\textsuperscript{N}}{}}\label{adaptN} In this section we turn our attention to the following question: 
can we remove all local traps from ADAPT-VQE without necessarily reaching ``controllability'' where
all local minima are exact?
It follows from Ref. \citenum{larocca_theory_2021} that one can systematically overparametrize a given ansatz of the form $\ket{\Psi} = e^{\theta_1 A_1}e^{\theta_2 A_2}\dots e^{\theta_M A_M}\ket{\phi_0}$ by simply repeating all the operators $N$ times
such that the number of parameters approaches or exceeds the dimension of the associated DLA. This ansatz repetition can be defined as:
\begin{equation}\label{dla2}
\ket{\Psi} = \prod_{i=1}^N\left(\prod_{j=1}^M e^{\theta_j^i A_j}\right)\ket{\phi_0}.
\end{equation}
We use the superscript on $\theta^i_j$ to indicate the $i$th independent parameter associated with the same operator $A_j$.  Because the same generators are being used, the DLA remains the same, but the number of parameters has now been multiplied by $N$.  We apply this idea to ADAPT-VQE in what we call ADAPT$^N$.  As with normal ADAPT, we begin with a reference $\ket{\phi_0}$.  We choose the first operator $A_1$ to add as before, but apply it to the reference $N$ times with $N$ distinct parameters: \ref{adaptN1}.
\begin{equation}\label{adaptN1}
\ket{\Psi_1} = e^{\theta_1^1 A_1}e^{\theta_1^2 A_1}\dots e^{\theta_1^N A_1}\ket{\phi_0}.
\end{equation}
The energy of $\ket{\Psi_1}$ is variationally minimized with respect to $\boldsymbol\theta$, beginning from $\boldsymbol\theta = \mathbf{0}$.  For each operator $A_i$ in the operator pool, we now measure the energy gradient associated with adding it to the front of the ansatz with a parameter of 0.
As with the original ADAPT-VQE, we choose the next operator in our ansatz, $A_2$, to be the operator with largest gradient magnitude.  We then add the operator in a collated fashion:
\begin{equation}\label{adaptN2}
\ket{\Psi_2} = \left(e^{\theta_2^1A_2}e^{\theta_1^1 A_1}\right)\left(e^{\theta_2^2A_2}e^{\theta_1^2 A_1}\right)\dots \left(e^{\theta_2^N A_2}e^{\theta_1^N A_1}\right)\ket{\phi_0}.
\end{equation}
This procedure of adding a new operator to the ``core'' ansatz, and replicating $N$ times, is then repeated until convergence. 
The intuitive way to understand ADAPT$^N$ is that ADAPT-VQE is adding $N$ parameters at a time, rather than the usual 1.  However, unlike the ``batched'' approach of Ref. \citenum{sapova_variational_2021}, our method does not add $N$ \textit{different} operators, a choice made to ensure that the DLA dimension does not increase with $N$. Because the ratio of the number of parameters to DLA dimension increases with $N$, we hypothesize that ADAPT$^N$ should exhibit fewer local minima with increased $N$.  

\begin{figure*}
\includegraphics[width=1\textwidth]{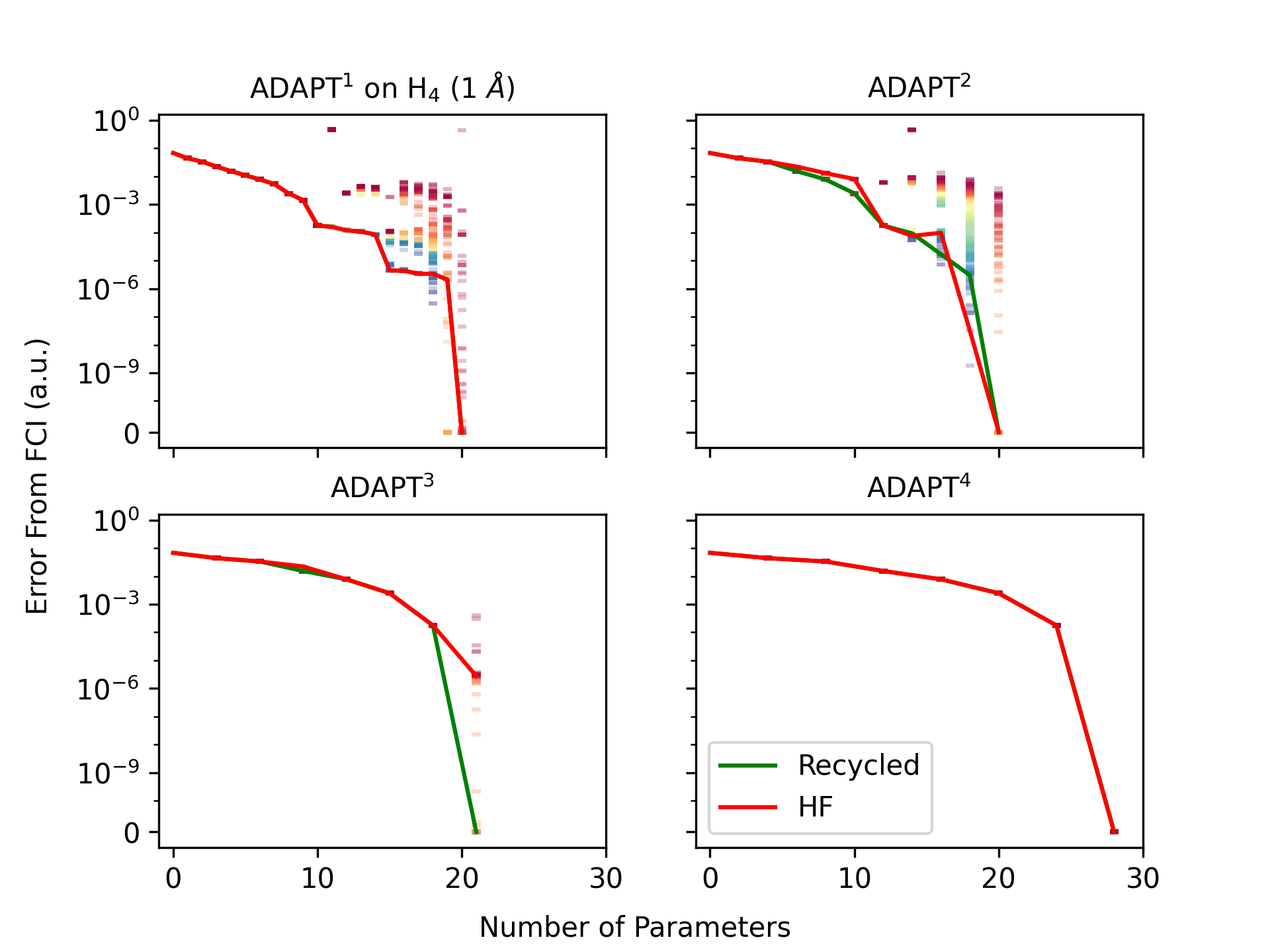}
\caption{ADAPT$^N$ for $N = 1,2,3,4$ on H$_4$ at 1 \AA{} separation.  The axes and colors are as in Fig. \ref{h4_1A_adapt}}\label{breadapt}  
\end{figure*}

To test this hypothesis, we consider first the numerically challenging linear H$_4$ at 1{\AA} bond distance in the STO-3G basis. 
In Figure \ref{breadapt}, we show the convergence of ADAPT$^N$ for $N\leq 4$ as a function of the number of parameters.  Like before, we consider 300 random initializations at each ansatz length in addition to the recycled initialization from the previous VQE subroutine, and the HF initialization.  We then plot the optimized results of all 302 initializations.  
For ADAPT$^1$ (i.e. standard ADAPT) and ADAPT$^2$, we see similar behavior.  Random guesses are frequently better than recycled ones, with random initializations of $\boldsymbol\theta$ landing in numerous local minima.  For ADAPT$^1$ and ADAPT$^2$, the landscape is dominated by local minima except when there are very few operators or when FCI has almost been reached.  For ADAPT$^3$, most of the local minima seem to be gone, and local minima never outperform the recycled initialization.  For ADAPT$^4$, every VQE subroutine appears to be free of local minima.  This gives us a simple way to make ADAPT-VQE (or similar VQEs) trap-free.  The downside is that more parameters and deeper circuits are required to introduce the same number of distinct operators.  Based on these simulations, the lower values of $N$ seem to be more depth-efficient at achieving any given accuracy at any given circuit depth, even if only the recycled initialization is tried.  Furthermore, based on Ref. \citenum{larocca_theory_2021}, the necessary $N$ to keep ADAPT-VQE trap-free will increase with the number of operators needed to describe a system, which will generally grow with system size.  We therefore expect this overparametrization strategy to scale poorly.  Contrary to the prediction of Ref. \citenum{wierichs_avoiding_2020}, however, we have shown that an ADAPT-like method can avoid local traps via overparametrization.

\section{Difficulty of Exhaustive Sampling}\label{exhaustive}
\begin{figure}[hbt!]
\includegraphics[width=.48\textwidth]{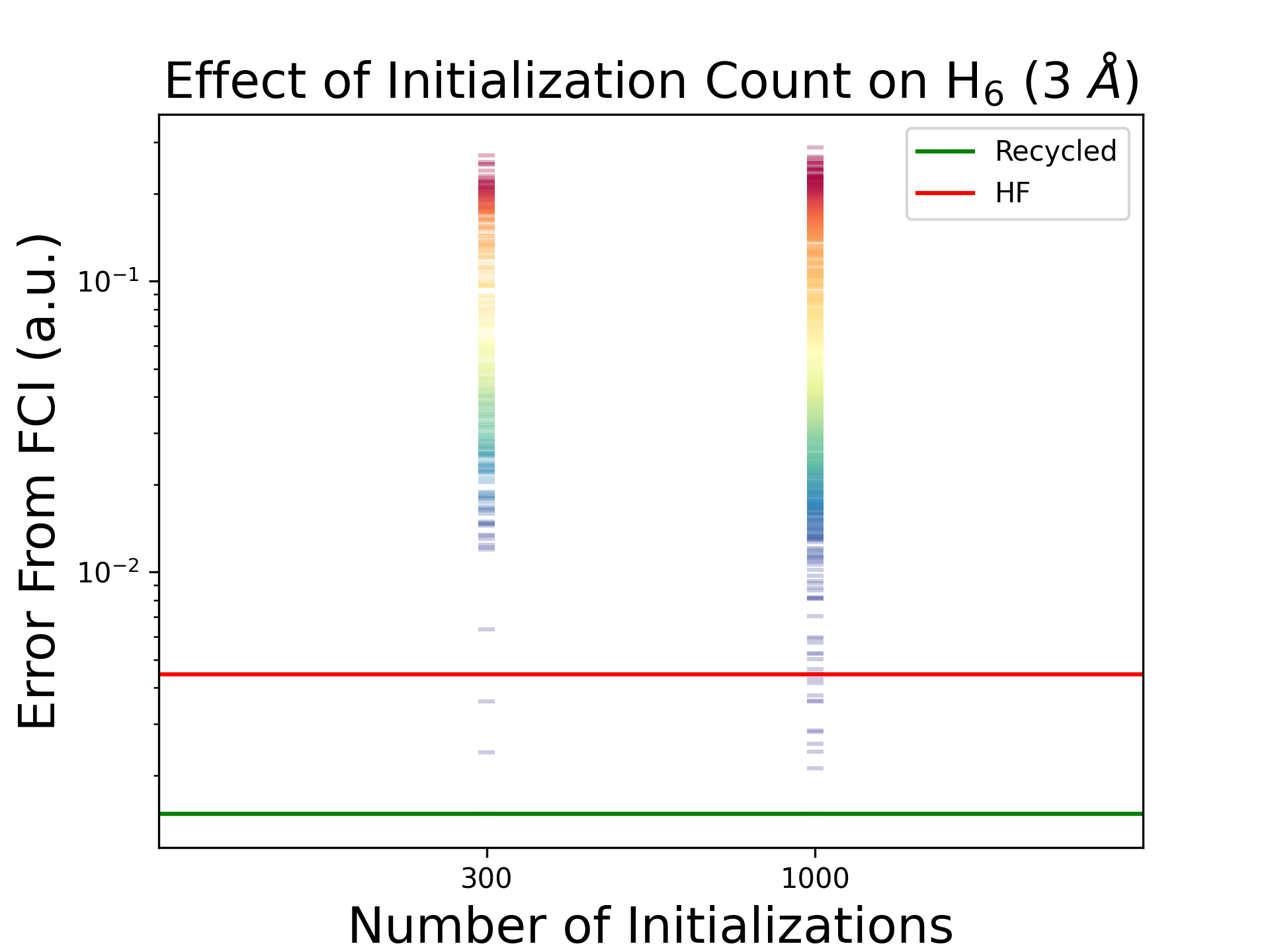}
\caption{Results of various initializations for the 60-parameter ADAPT-VQE ansatz for H$_6$ at 3\AA.  The original 300 random initializations from Figure \ref{h6_3A_adapt} are shown on the left.  1000 random initializations are depicted on the right.}\label{h6_3A_1000_adapt}  
\end{figure}

While sampling over parameter initializations provides a path for enumerating different local traps in the landscape, 
the computational cost of such sampling is high, 
and any such study will necessarily exhibit artifacts arising from insufficient sampling. 
While our best efforts were made to avoid such artifacts,
we have noticed a few.
Take for example, Figure \ref{h6_3A_adapt}, where the best solution at 60 parameters is very slightly worse than the best solution at 59 parameters.  We tried using 1000 random initializations at 60 parameters (Fig. \ref{h6_3A_1000_adapt}), and were still unable to improve on the recycled initialization at 60 parameters, let alone the best initialization at 59 parameters. 
While these small artifacts persist, 
the general trends seem converged. 
For instance, in Fig. \ref{h6_3A_1000_adapt}, 
increasing the number of initializations increases the number of solutions found. 
The distribution of the solutions is essentially unchanged,
with the median values (yellow-green) staying roughly in the same place. 

\begin{figure}
\includegraphics[width=.48\textwidth]{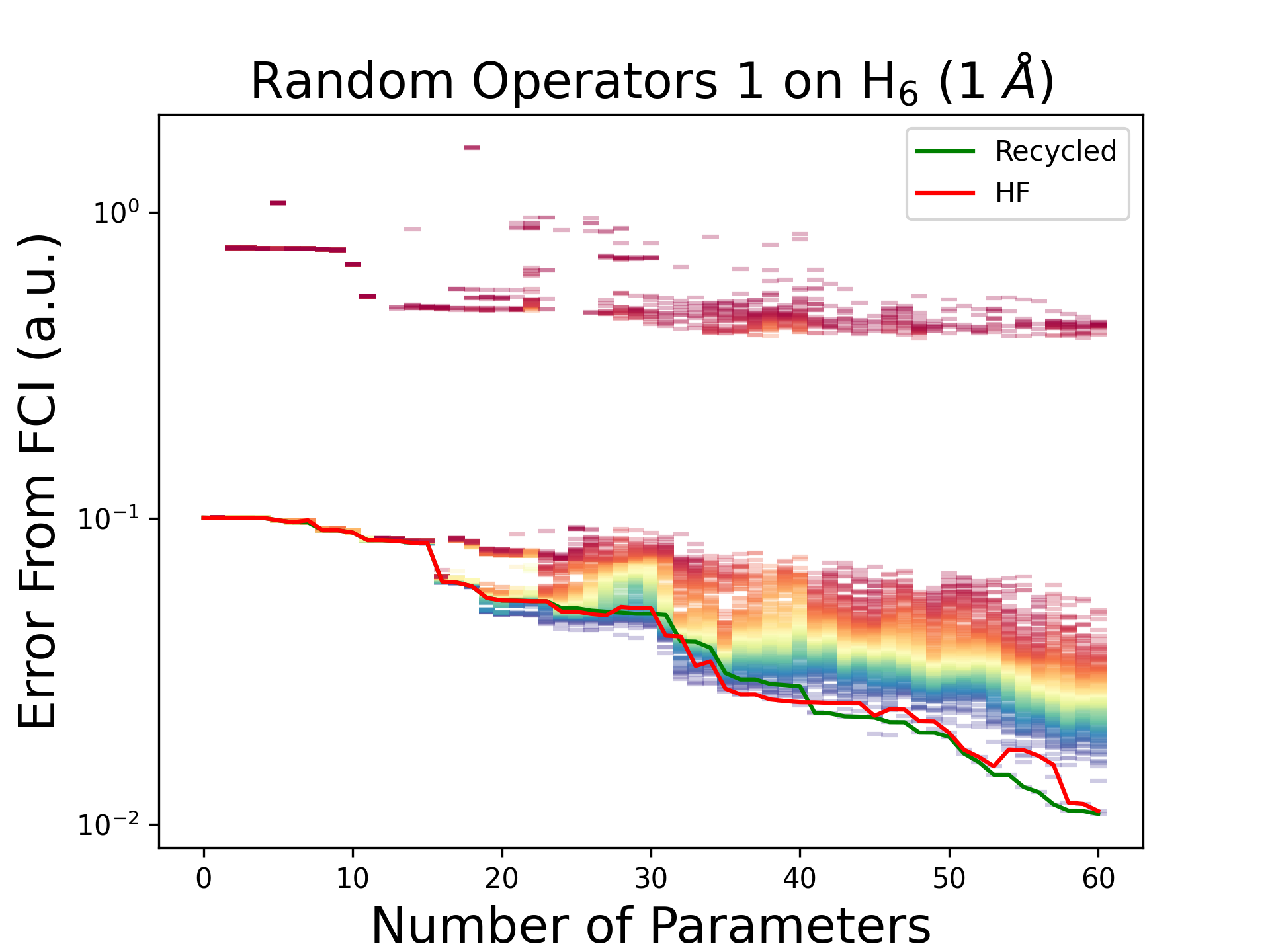}
\caption{Random Reordering 1 of the 71 operators identified in Fig. \ref{h6_1A_adapt}.  (Computer time was exhausted before all 71 operators were added in this case.)  The $x$-axis corresponds to the number of operators in the ansatz at a given step.  The $y$-axis corresponds to the error from the exact FCI energy.  The red curve (HF) corresponds to the energy obtained through BFGS minimization using an all-zero initialization.  The green curve corresponds to the energy obtained through BFGS minimization using the standard ADAPT-VQE recycled initialization.  The colored dashes correspond to all the energies obtained through BFGS optimizations, with red being the highest energy and violet the lowest.}\label{h6_1A_reorder_1}
\end{figure}

\begin{figure}
\includegraphics[width=.48\textwidth]{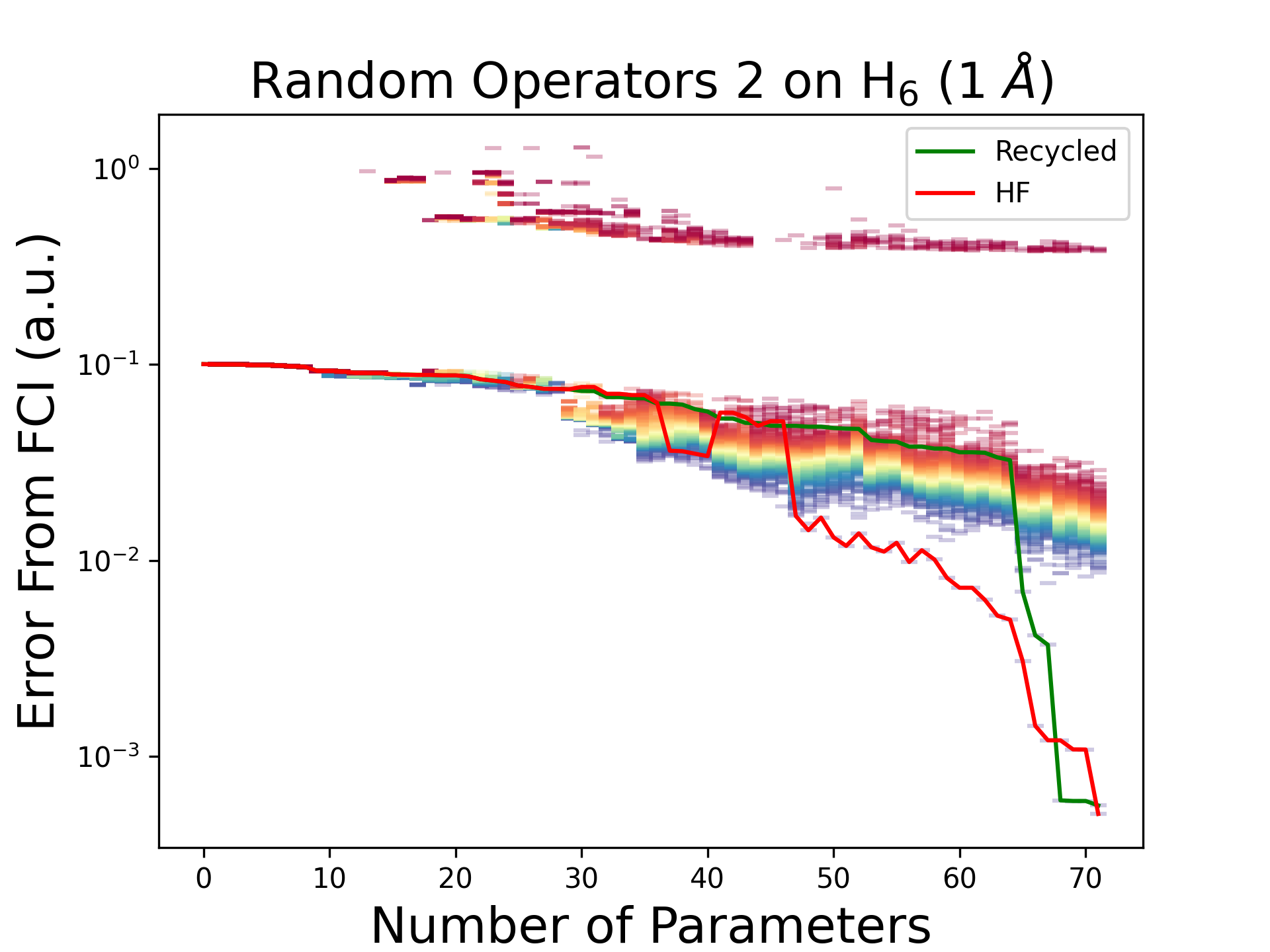}
\caption{Random Reordering 2 of the 71 operators identified in Fig. \ref{h6_1A_adapt}.  Labels same as in Fig. \ref{h6_1A_reorder_1}}\label{h6_1A_reorder_2}
\end{figure}
\begin{figure}
\includegraphics[width=.48\textwidth]{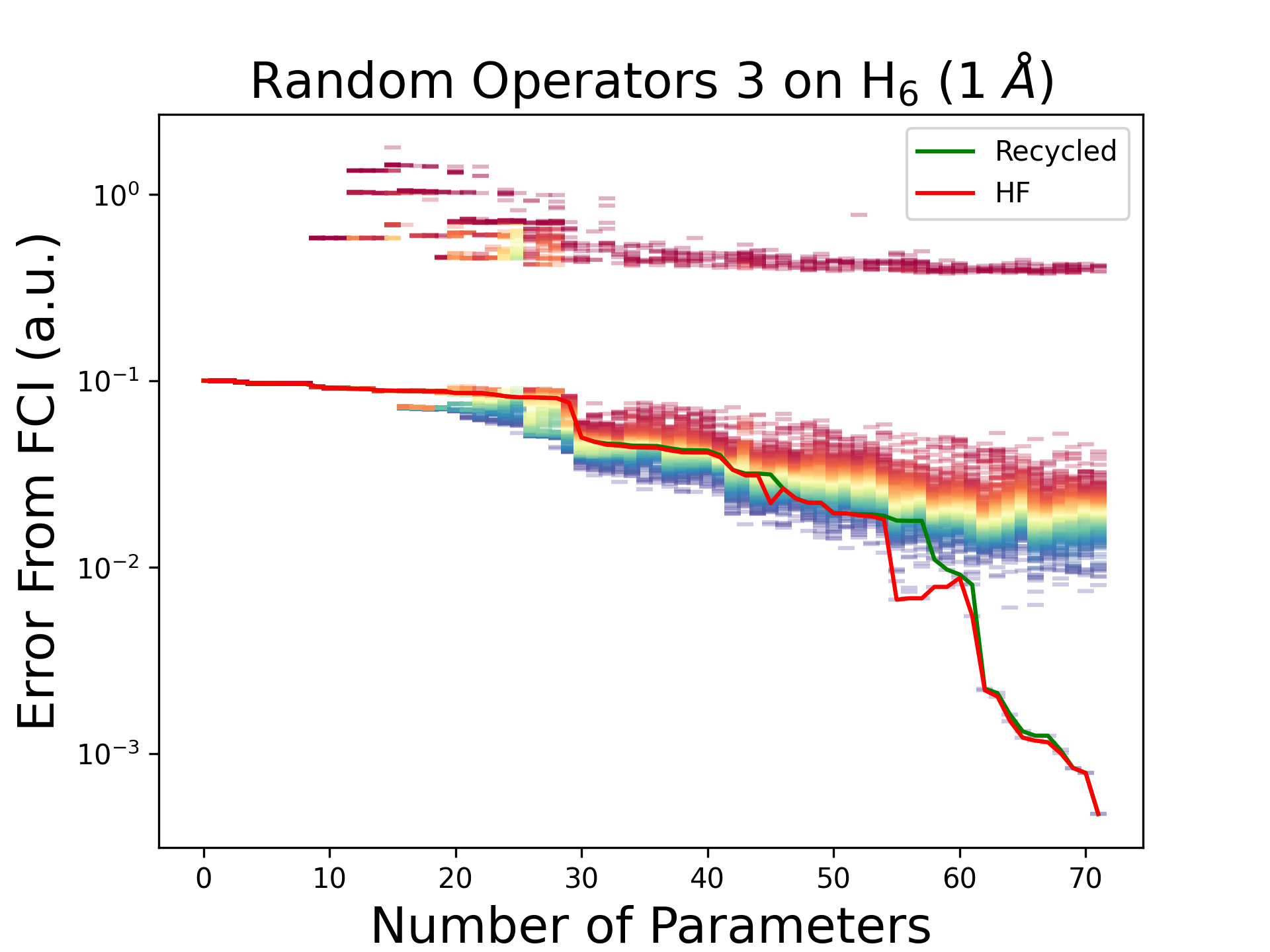}
\caption{Random Reordering 3 of the 71 operators identified in Fig. \ref{h6_1A_adapt}.  Labels same as in Fig. \ref{h6_1A_reorder_1}}\label{h6_1A_reorder_3}
\end{figure}

\section{Importance of gradient-guided ansatz}\label{sophia}
In Figs. \ref{h4_1A_adapt}-\ref{beh2_adapt}, we demonstrated that the recycled (and HF) initialization performed far better than the average local minimum. 
In order to ascertain the role that the ADAPT-VQE ansatz construction has on the ability of the ansatz to burrow, we carry out similar simulations to those presented in Fig. \ref{h6_1A_adapt}, 
except that before performing the VQE optimizations, we shuffle the operators in the ansatz. 
As a result, the ansatz is no longer constructed by ADAPT-VQE. 
The results are shown in Figs. \ref{h6_1A_reorder_1}-\ref{h6_1A_reorder_3}. In each of the 3 random reorderings, 
we find that the large gap that appeared in Fig. \ref{h6_1A_adapt} between the recycled (and HF) minimum and the local traps, largely disappears when considering a randomly shuffled ansatz. 
This is clear evidence that it is not sufficient to only slowly optimize the parameters one-at-a-time via recycled initialization, 
but that the ansatz must be grown in an efficient manner as well if one is to observe the full burrowing effect.

\end{document}